\documentclass[12pt]{article}
\input epsf
\usepackage{amssymb}
\usepackage{amsmath}

\usepackage{boxedminipage}
\usepackage{listings}
\usepackage{minitoc}

\makeatletter
\@addtoreset{equation}{section}
\makeatother

\def\be{\begin{equation}}
\def\ee{\end{equation}}

\def\ba{\begin{array}}
\def\ea{\end{array}}

\newcommand{\bea}{\begin{eqnarray}}
\newcommand{\eea}{\end{eqnarray}}

\def\K{K{\"a}hler}

\begin{document}
\hfill{}

\begin{flushright}
CERN-PH-TH/2011-242\\
SU-ITP-2011-48\\
\end{flushright}
\vspace{.5cm}
\vskip 1cm

\vspace{10pt}

\begin{center}
{ \large { \bf  Creation of Matter in the  Universe   \\

\vskip 0.5 cm

and Groups of Type E7}}

\

{\bf  Sergio Ferrara$^{1}$ and Renata Kallosh$^2$
 } \\[7mm]

{\small

$^1$ Physics Department, Theory Unit, CERN, 1211 Geneva 23, Switzerland and\\
\vspace{6pt}
INFN, Laboratori Nazionali di Frascati, Via Enrico Fermi, 40,  00044 Frascati, Italy and \\
\vspace{6pt}
Department of Physics and Astronomy, University of California, Los Angeles, CA 90095-1547,USA\\
\vspace{6pt}
$^2$ Physics Department, Stanford University, Stanford   CA 94305-4060, USA\\ }
\vspace{6pt}

\vspace{10pt}

\vspace{24pt}

\underline{ABSTRACT}

\end{center}

We  relate the mechanism of matter creation in the universe after inflation to a simple and universal mathematical property of extended $N> 1$  supergravities and related compactifications of superstring theory. We show that in all such models, the inflaton field may decay into vector fields due to a nonminimal scalar-vector coupling. This coupling is compulsory for all scalars except  $N=2$ hyperscalars. The  proof is based on the fact that all extended supergravities described by symmetric coset spaces ${G\over H}$  have  duality groups $G$ of type E7, with exception of  $U(p,n)$ models.  For $N=2$ we prove separately that special geometry requires a non-minimal scalar-vector coupling. Upon truncation to $N=1$ supergravity, extended models generically preserve the non-minimal scalar-vector coupling, with exception of $U(p,n)$ models and hyperscalars.  
For some string theory/supergravity inflationary models, this coupling provides the only way to complete the process of creation of matter in the early universe.

\vfill

\newpage


\

\section{Introduction}

According to inflationary theory, all elementary particles populating our universe have been created as a result of the process of preheating and reheating of the universe after inflation. During inflation, the energy density of the universe is dominated by the energy of the inflaton field $\phi$. At the end of inflation, the inflaton field may rapidly transfer a part of its energy to other particles and fields in the process of preheating, a non-perturbative process which may occur due to a combination of parametric resonance, tachyonic instability, and rescattering of produced particles and waves. However, this process does not take away all energy of the inflaton filed, so it should be followed by reheating, a perturbative particle production and the inflaton decay  \cite{Kofman:1994rk}. Reheating leads to a complete decay of the inflaton field only if the inflaton field can actually decay to other particles, i.e. if the theory allows the process  $\phi \to {\it anything}$  rather than some interaction $\phi + \phi \to {\it anything}$. There are some theories where the decay $\phi \to {\it anything}$ is forbidden, reheating is incomplete, and the universe eventually becomes  dominated by the oscillating inflaton field, or by other scalar fields produced during preheating. Such theories are cosmologically unacceptable. Therefore it it important to understand whether there are some deep theoretical reasons to expect the existence of interactions which could lead to a complete decay of the inflaton field and of all other scalar fields which could be produced at the end of inflation. 

A complete decay of the scalar fields may occur due to the scalar-dependent vector coupling, which is possible in 
$N=1$ supergravity \cite{Cremmer:1982en}:
\begin{eqnarray}
& -\frac{1}{4} \Big ({\rm Re}\, f_{\alpha \beta}(\varphi)\Big )F^{\alpha}_{\mu \nu}F^{\beta \mu\nu}
+\frac{i}{4}\Big ({\rm Im}\, f_{\alpha \beta}(\varphi)\Big )F^{\alpha}_{\mu \nu}\tilde F^{\beta \mu\nu} \ .
\label{nonminimal}
 \end{eqnarray}
Here the function $ f_{\alpha \beta}(\varphi)$  is  holomorphic. However, whereas in $N= 1$ supergravity the dependence of $ f_{\alpha \beta}(\varphi)$ on scalars is possible, it is not required.
The preference  is often  given to the  minimal, scalar-independent vector couplings, where
\begin{eqnarray}
 {\rm Re}\, f_{\alpha \beta}(\varphi)= \delta_{\alpha \beta}\, , \qquad {\rm Im}\, f_{\alpha \beta}(\varphi)=0 \ .
 \label{minimal}
 \end{eqnarray}
Indeed, the Ockham's razor principle generally recommends, when faced with competing hypotheses that are equal in other respects, selecting the one that makes the fewest new assumptions. At the level of phenomenological $N=1$ supergravity the non-minimal vector coupling requires an unmotivated extra assumption about the function $f_{\alpha \beta}(\varphi)$ and should therefore be avoided, unless such a motivation is provided. Moreover, in some textbooks $N=1$ supergravity is presented only in the form  (\ref{minimal}), see e.g. \cite{Wess:1992cp}.

It has been   recently explained in \cite{Kallosh:2011qk} that in some models of inflation in supergravity which provide an arbitrary inflaton potential  \cite{Kallosh:2010ug} there is no decay route for the inflaton field, unless one  involves the non-minimal vector coupling. An analogous situation with reheating was noticed earlier in  the string theory modular inflation models \cite{BlancoPillado:2004ns,Barnaby:2009wr} as well as in related supergravity inflation models \cite{Endo:2006qk}. In \cite{BlancoPillado:2004ns,Barnaby:2009wr} it was explained that in string theory standard model particles can live on wrapped D7 branes. In such case
the inflaton which is the combination of the 4-cycle volume modulus $Y$ and its axionic partner $X$ naturally couple to vectors
\be
Y F_{\mu\nu}  F^{\mu\nu}\, , \qquad X F_{\mu\nu}  \tilde F^{\mu\nu} \, .
\ee
This provides a possibility of the creation of matter after inflation in these models via the inflaton decay into vectors.
However, it was not clear  whether this is just a specific example of the non-minimal vector coupling, or a generic feature of all  phenomenological $N=1$ supergravity models derived from string theory via compactification and/or  from extended supergravities. Since this issue appears in many  inflationary models based on supergravity, we decided to investigate it.

We would like to note here that the non-minimal coupling of vectors is always complemented by the Pauli couplings of the form \cite{Cremmer:1982en}
\be
{\partial   \, f_{\alpha \beta}(\varphi)\over \partial  \varphi^i} \, \bar \chi ^i \sigma_{\mu\nu} \lambda ^\alpha F^{\beta \mu\nu} \ ,
\label{Pauli}\ee
which allow the decay of the fermions $\bar \chi^i$  into  gaugino $\lambda ^\alpha$ and a vector $F^{\beta \mu\nu}$.  This is in addition to the usual gaugino-gravitino-vector Pauli terms. Therefore if the superheavy scalar fields in the inflaton multiplet may completely decay after inflation, the same conclusion will be valid for their fermionic superpartners.

The purpose of this note is to prove that the  non-minimal scalar-dependent vector coupling is compulsory (except for hyperscalars) in all $N>1$ supergravities, that have scalars,  and in related  superstring theory compactifications.  This provides a motivation to use such couplings  in phenomenological $N=1$ models inspired by the superstring theory/extended supergravity. We will also prove that generic  $N>1$ supergravities  consistently truncated to $N=1$ supergravity forbid the minimal choice (\ref{minimal}). The only exception are $U(p,n)$ models and the ones where scalars originate from $N=2$ hypers.

\section{Extended d=4 $N>1$ supergravity}

The bosonic part of $d=4$  $N>1$  supergravities depends on metric, vectors and scalars. In particular, the action depends on  Abelian vectors ${\cal A}_\mu^{\Lambda}$ via the field strength $F_{\mu\nu}^{\Lambda}= \partial_\mu {\cal A}_\nu^{\Lambda}-\partial_\nu {\cal A}_\mu^{\Lambda}$, on scalars and on the metric\footnote{In $N=2$ models the kinetic term for (vector multiplet) scalars has a \K\, form: $g_{i\bar j} \partial_\mu \phi^i  \partial^\mu \bar \phi^{\bar j}$ with $g_{i\bar j}= \partial_i \partial_{\bar j} K$ and $K$ further restricted in virtue of eq. (\ref{Riem}), see below.}  \
\be
S_{\rm cl}(F, \phi, g)= {1\over 4 \kappa^2} \int d^4x \, e \Big( -{1\over 2} R+ {\rm Im} {\cal N}_{\Lambda \Sigma}
F_{\mu\nu}^\Lambda F^{\mu\nu \Sigma}  + {1\over 2 \, e} {\rm Re} {\cal N}_{\Lambda \Sigma} \epsilon^{\mu\nu\rho\sigma} F_{\mu\nu}^\Lambda F_{\rho\sigma}^\Sigma + {1\over 2} g_{ij}(\phi) \partial_\mu \phi^i \partial^\mu \phi^j \Big )
\label{0}\ee
Here the kinetic term for vectors ${\cal N}_{\Lambda \Sigma}(\phi)$ in general depends on scalars. The matrix ${\rm Im} {\cal N}_{\Lambda \Sigma}
$ is a metric in the vector moduli space.  It must be negative definite to provide the positive energy and it must be invertible so that a consistent quantization is possible. These properties of the kinetic matrix ${\cal N}_{\Lambda \Sigma}(\phi)$ will be used in the following.

One should keep in mind that the $N=1$ supergravity vector coupling $f_{\alpha\beta}(z)$ corresponds to $-4 i \,\overline{\cal N}_{\Lambda \Sigma}$. The manifold of scalars  for $3\leq N\leq 8$ is in each case a unique  symmetric coset space ${G/H}$, where the group $G$ is the Gaillard-Zumino duality symmetry \cite{Gaillard:1981rj}, see Table 1.
In case $N=2$ it can be a symmetric coset space ${G/H}$, see Table 2 for 7 choices,  or it can be also a non-symmetric space described by a Hodge \K \, manifold of $N=2$ special geometry (we will discuss separately the case of $N=2$ hypermultiplets, which are decoupled from the vector multiplets) \cite{deWit:1984pk}-\cite{Ceresole:1995ca}.

\begin{table}[t]
\begin{center}
\begin{tabular}{|c||c|c|}
\hline
$N$& $
\begin{array}{c}
\\
$G$ \\
~
\end{array}
$ & $
\begin{array}{c}
\\
  \mathbf{ R}
\\
~
\end{array}

$ \\ \hline\hline
$
\begin{array}{c}
\\
N=3 \\
~
\end{array}
$ & $U(3,n)$ & $ \mathbf{(3+n)}$     \\ \hline
$
\begin{array}{c}
\\
N=4 \\
~
\end{array}
$ & $SL(2, \mathbb{R})\otimes {SO(6,n)}$ & $\mathbf{(2, 6+n)}$   \\ \hline
$
\begin{array}{c}
\\
N=5 \\
~
\end{array}
$ & $SU(1,5)$ & $ \mathbf{ 20}$   \\ \hline
$
\begin{array}{c}
\\
N=6 \\
~
\end{array}
$ & $SO^{\ast }(12)$ & $\mathbf{ 32}$
 \\ \hline
$
\begin{array}{c}
\\
N=8 \\
~
\end{array}
$ & $E_{7\left( 7\right) }$ & $\mathbf{ 56}$   \\ \hline
\end{tabular}
\end{center}
\caption{ $N\geqslant 3$ supergravity sequence of groups $G$ of  the corresponding ${G\over H}$ symmetric spaces, and their symplectic representations  $\mathbf{R}$}
\end{table}

\begin{table}[t]
\begin{center}
\begin{tabular}{|c||c|}
\hline
$
\begin{array}{c}
\\
$G$\\
~
\end{array}
$ & $\mathbb{\mathbb{}}
\begin{array}{c}
\\
  \mathbf{ R}\\
~
\end{array}
$ \\ \hline\hline
$
\begin{array}{c}
\\
{U(1,n)}
\end{array}
$ & $
\begin{array}{c}

\mathbf{(1+n)_c}\\

\end{array}
$ \\ \hline
$
\begin{array}{c}
\\
{SL(2, \mathbb{R})}\otimes SO(2,n)
~ ~

\end{array}
$ & $
\begin{array}{c}
\mathbf{(2, 2+n)}

\end{array}
$ \\ \hline
$
\begin{array}{c}
\\
SL(2, \mathbb{R})
~
\end{array}
$ & $
\begin{array}{c}
\\
\mathbf{4}
\end{array}
$ \\ \hline
$
\begin{array}{c}
\\
Sp(6,\mathbb{R})~
\end{array}
$ & $
\begin{array}{c}
\\
\mathbf{14}' \\

\end{array}
$ \\ \hline
$
\begin{array}{c}
\\
SU(3,3)\end{array}
$ & $
\begin{array}{c}
\\
\mathbf{20}
~
\end{array}
$ \\ \hline
$
\begin{array}{c}
\\
SO^{\ast }(12)~
\end{array}
$ & $
\begin{array}{c}
\\
\mathbf{32}
~
\end{array}
$ \\ \hline
$
\begin{array}{c}
\\
E_{7\left( -25\right) }
~
\end{array}
$ & $
\begin{array}{c}
\\
\mathbf{56}
~
\end{array}
$ \\ \hline
\end{tabular}
\end{center}
\caption{$N=2$ choices of groups $G$ of the  ${G\over H}$ symmetric spaces and their symplectic representations  $\mathbf{R}$. The last four lines refer to ``magic $N=2$ supergravities''.}
\end{table}

The first set of all extended supergravities $d=4$  $N>1$ based on  symmetric coset spaces ${G/H}$ has a remarkable property that almost all groups $G$ are of type E7 \cite{Brown,Borsten:2009zy}, see Tables 1, 2. This universal property will be used to prove that for all extended supergravities $d=4$  $N>1$ based on  symmetric coset space ${G/H}$ the vector coupling ${\cal N}_{\Lambda \Sigma}(\phi)$ must be scalar dependent, with exception of hyper scalars. In this first set
\be
{\partial \over \partial \phi ^i } \, {\cal N}_{\Lambda \Sigma} (\phi)\equiv \partial_i   \, {\cal N}_{\Lambda \Sigma}\neq 0\, .
\label{NM}\ee
In the remaining non-symmetric  $N= 2$ supergravities we will prove that eq. (\ref{NM}) is required  for a consistency of the special Hodge \K \, manifold.  It is quite remarkable that when $\partial_i   \, {\cal N}_{\Lambda \Sigma}\neq 0$ $N>1$ supergravities require a Pauli coupling of the type (\ref{Pauli}). We will present the details in the $N=2$ case below.

In the cosmological context of creation of matter in the early universe it is important that groups of type E7 do not admit a quadratic bilinear polynomial.

\subsection{Groups of type E7 and Compulsory Non-minimal Vector Coupling in $N>1$}

Simply put, groups of type E7 have a symplectic representation $\mathbf{R }$ admitting a symmetric quartic invariant polynomial, but not a quadratic one.   We will elaborate further with relevant examples.   These were first defined and studied in \cite{Brown}.  This was about a decade before\footnote{ The first paper in \cite{Brown} by Brown was submitted in 1967 and published in 1969.} the discovery of  supergravity.   In fact, surprisingly,  all  extended supergravities $2 \leq N\leq 8$ described by coset spaces ${G\over H}$  have  $G$ of type E7
\cite{Borsten:2009zy,Ferrara:2011gv}, with the exception of  $N=2$ group $G= U(1,n)$ and $N=3$   group
$G= U(3,n)$. These $G= U(p,n)$ groups are not of type E7, since they have a primitive quadratic symmetric invariant (in addition to a symplectic bilinear form).

For $N$ between 8 and 3 there are 4 classes of type E7 groups:
\be
E_{7\left( 7\right) }\, ,  \quad  SO^{\ast }(12)\, ,  \quad SU(1,5)\, ,  \quad SL(2, \mathbb{R})\otimes {SO(6,n)}\, ,
\ee
and one class $U(3,n)$ which is not of type E7,  see Table 1.
 In $N=2$ cases of symmetric spaces there are 6 classes of  type E7 groups \cite{Gunaydin:1983rk}, see Table 2
\be
E_{7\left( -25\right) }\, ,   \quad SO^{\ast }(12)\, ,   \quad SU(3,3)\, ,   \quad Sp(6,\mathbb{R})\, ,   \quad SL(2, \mathbb{R})\, ,
 \quad   {SL(2, \mathbb{R})}\otimes SO(2,n)\, ,
\ee
and one class ${U(1,n)} $ which is not of type E7.
Here $n$ is the integer describing  the number of matter multiplets for $N=4, 3, 2 $.

Given a symplectic representation $\mathbf{R }$ of dimension $r$
\be
(\mathbf{R }_1, \mathbf{R }_2)= - (\mathbf{R }_2, \mathbf{R }_1)
\ee
one can construct a quartic invariant
$
q(\mathbf{R }_1, \mathbf{R }_2, \mathbf{R }_3, \mathbf{R }_4)
$
which is completely symmetric so that the quartic $G$-invariant polynomial is
\be
{\cal I}_4(\mathbf{R })= q(\mathbf{R }_1, \mathbf{R }_2, \mathbf{R }_3, \mathbf{R }_4)_{\mathbf{R }_1=\mathbf{R }_2=\mathbf{R }_3=\mathbf{R }_4=\mathbf{R }}\, .
\ee
The famous example of such a quartic invariant in $G=E_{7(7)}$  is the Cartan-Cremmer-Julia invariant \cite{Cartan} which defines the area of the $N=8$ black hole horizon \cite{Kallosh:1996uy}.
In groups of type E7 the quadratic symmetric invariant  is generically not available, only an antisymmetric one is available, therefore
\be
(\mathbf{R }_1, \mathbf{R }_2)_{\mathbf{R }_1=\mathbf{R }_2}=0\, .
\ee
Cases of  $N=2$ with $G= U(1,n)$ and $N=3$ with $G= U(3,n)$ in fact have a quadratic invariant hermitian form $(\mathbf{R }_1, \overline {\mathbf{R }}_2)$ whose imaginary part is the symplectic (antisymmetric) invariant and whose real part is the symmetric quadratic invariant defined as follows (see Appendix)
\be
{\rm Re}\,  (\mathbf{R }_1, \overline {\mathbf{R }}_2)_{\mathbf{R }_1=\mathbf{R }_2}={\cal I}_2(\mathbf{R })\equiv  \sqrt {|{\cal I}_4(\mathbf{R })|}\neq 0\, ,
\ee
\be
{\rm Im}\,  (\mathbf{R }_1, \overline {\mathbf{R }}_2)_{\mathbf{R }_1=\mathbf{R }_2}=0\, .
\ee
Thus, the fundamental representations of pseudounitary groups U(p,n), which have a hermitean quadratic invariant form,  do not strictly qualify for groups of type E7. The reason is that  they have, in addition to the symplectic invariant, also a symmetric real quadratic invariant instead of a quartic one.  This is the reason why from all models of extended supergravities only  $G= U(p,n)$  with $p=1,3$  are not groups of type E7.

Note,
however, in all cases with $n\neq 0$ and $N=2,3$ these groups are non-compact and the relevant quadratic invariants can not describe the negative definite vector couplings, as we will see later. In $n=0$ case the  groups are compact but there are no scalars. All of this will be important for our consequent analysis. In particular, these  $U(p,n)$ models will form an exceptional case in truncation to $N=1$.

We will now prove here that all vector couplings in extended supergravities with scalars must be non-minimal: they must depend on all scalars with the exception of $N=2$ hypers. For type E7 models this follows from  the absence of a symmetric quadratic invariant tensor. For symmetric spaces with $G=U(p,n)$  it follows from the fact that  their symmetric quadratic invariant tensor is not negative definite.

The group $G$ is first embedded into an $Sp(2n_v, \mathbb{R})$ one. The duality symmetry of the theory transforms the symplectic representation as follows
\be
\mathbf{R }'= {\mathcal{S}} \mathbf{R } \ ,
\ee
where a real symplectic $Sp(2n_v, \mathbb{R})$ matrix is
\be \label{Symplectic}
{\mathcal{S}}= \begin{pmatrix}
  A & B \\
  C & D
\end{pmatrix}\ , \qquad \mathcal{S}^t \;   \Omega \; \mathcal{S}= \Omega\ , \qquad \Omega = \begin{pmatrix}
 0 & -\mathbb{I} \\
 \mathbb{I} & 0 
\end{pmatrix}\ ,
 \ee
so that
$
A^t C- C^t A=0$, ~$ B^t D- D^t B=0$, ~$A^t D - C^t B=1
$. The gauge kinetic term ${\cal N}$ transforms via fractional transformations
\be
{\cal N}'=( C+  D {\cal N})(  A +  B {\cal N})^{-1}\, .
\label{sctransf}\ee
A new symmetric symplectic matrix of dimension $2n_v\times 2n_v$ can be constructed from the $n_v\times n_v$ matrix ${\cal N}$ as follows \cite{Breitenlohner:1987dg}
  
\be
\mathcal{M}\, (\mathcal{N})=   \begin{pmatrix}
 {\rm Im} \,  \mathcal{N} + {\rm Re}\mathcal{N} \, ({\rm Im} \mathcal{N})^{-1} \, {\rm Re}\mathcal{N}   & -{\rm Re}\,  \mathcal{N} ({\rm Im} \,  \mathcal{N})^{-1} \\
 \\
-({\rm Im}   \mathcal{N})^{-1}\,  {\rm Re}\mathcal{N}\, &( {\rm Im} \,  \mathcal{N})^{-1}
\end{pmatrix}
\ee
such that
\be
\mathcal{M}^T \Omega \mathcal{M}=\Omega\, , \qquad \mathcal{M}^T = \mathcal{M}\, .
\ee
This matrix transforms as a tensor under duality transformations. If ${\cal N}$ would be constant, $\mathcal{M}$ would also be constant. This would imply that there is a symmetric invariant tensor of the group $G$. However, none of these theories with E7 type $G$ has an invariant symmetric quadratic form for the symplectic representation $\mathbf{R }$ of $G$, by definition. The only extended supergravities, which have a symmetric quadratic invariant are the $N=3$ and $N=2$ theories with $U(p,n)$ groups.  However, their quadratic real form does not have all negative eigenvalues but has a Lorentzian structure, having the signature $(p,n)$,  due to the non-compactness of $G$ for $n\geq 1$: $U(3, n)$ for $N=3$ and $U(1, n)$ for $N=2$, see the first row in Table 1 and Table 2, respectively. Note that  ${\rm Im} \,  \mathcal{N}$ and $\mathcal{M}$ must be negative definite: for ${\rm Im} \,  \mathcal{N}$ this is a condition for positive energy, for $\mathcal{M}$ the proof can be found in \cite{Ceresole:1995ca}. Thus, only in absence of scalars when $n=0$ in $U(3, n)$ and $U(1, n)$ models the constant matrix $\mathcal{M}$ exist, but they are irrelevant since there are no scalars and the Gaillard-Zumino duality group \cite{Gaillard:1981rj} becomes a compact group $U(p)$.
This proves eq. (\ref{NM}) for all scalar-dependent $N>2$ supergravities based on coset spaces.

\subsection{$N=2$}

The case of $N=2$ requires a separate study since scalars may belong to either vector multiplets or hypermultiplets \cite{deWit:1984pk,Ceresole:1995ca}. Hypermultiplets and vector multiplets are decoupled: the interactions of hyper multiplets are described by a quaternionic geometry whereas the interactions of the vector multiplets are described by the special geometry. Therefore in $N>2$
\be
{\partial \over \partial \phi ^i } \, {\cal N}_{\Lambda \Sigma} (\phi)\neq 0 \,  \qquad {\rm  for \; all \; scalars} \ ,
\ee
 whereas in $N=2$ the kinetic term of vectors depends  on all scalars from the vector multiplets $\phi ^i_v$
and does not depend on any scalars from the hypermultiplets $q^u$
\be
{\partial \over \partial \phi ^i_v  } \, {\cal N}_{\Lambda \Sigma} (\phi)\neq 0 \, , \qquad {\partial \over \partial q^u  } \, {\cal N}_{\Lambda \Sigma} (\phi)= 0\, .
\ee

\subsubsection{ $N=2$ special geometry and Compulsory Non-minimal Vector Coupling}
In $N=2$ special geometry \cite{deWit:1984pk}-\cite{Ceresole:1995ca}
 the Riemann tensor of the \K\, manifold is
\be
R_{i\bar j k\bar l}=- (g_{i\bar j } g_{k\bar l }+  g_{i\bar l } g_{k\bar j }) + C_{ik p} C_{\bar j\bar l \bar p} g^{p\bar p} \ ,
\label{Riem}\ee
where $C_{ijk}$ is a covariantly holomorphic 3-form defined on a Hodge \K\, manifold.  The theory is based on  covariantly holomorphic symplectic section $(L^\Lambda, M_\Lambda)$.  We also define
 $f_i^\Lambda = D_i L^\Lambda$, $h_{i\Lambda} = D_i M_\Lambda$.  The \K\, metric in these notation is
\be
g_{i\bar j}= -2  {\rm Im} \,  \mathcal{N}_{\Lambda \Sigma} f_i^\Lambda \bar f_{\bar j}^\Sigma
\label{g}\ee
and the $C_{ikj}$ tensor is
\be
C_{ijk}= f_i^\Lambda \, \partial_j \, \overline{\mathcal{N}}_{\Lambda \Sigma} \, f_k^\Sigma\, .
\label{C}\ee
Here  $\mathcal{N}_{\Lambda \Sigma}$ is the vector kinetic matrix, providing the following relations:
 \be
 M_\Lambda=
 \mathcal{N}_{\Lambda \Sigma} L^\Sigma\, , \qquad  {\mathcal D}_{i}  M_\Lambda= \overline { \mathcal{N}}_{\Lambda \Sigma} {\mathcal D}_{i}  L^\Sigma\, .
 \ee
 It follows that
\be
\partial_i   \, {\cal N}_{\Lambda \Sigma} L^\Sigma =- ({\cal N}- \overline{{\cal N}})_{\Lambda \Sigma} f^\Sigma _i\, .
\label{dN}\ee
The Pauli terms with non-minimal vector couplings in $N=2$ supergravity are
\be
C_{ijk}\bar \lambda^{iA} \sigma_{\mu\nu} \lambda^{jB} F^{k \mu\nu}\epsilon_{AB}=   \partial_j \, \overline{\mathcal{N}}_{\Lambda \Sigma} \,  \bar \lambda^{\Lambda A} \sigma_{\mu\nu} \lambda^{i B} F^{\Sigma \mu\nu} \epsilon_{AB}
\ee
and for non-vanishing $\partial_i \, \overline{\mathcal{N}}_{\Lambda \Sigma}$ they remain upon a consistent truncation in the corresponding version of $N=1$ supergravity in agreement with  eq. (\ref{Pauli}).

 For constant ${\cal N}_{\Lambda \Sigma}$ eq. (\ref{dN})  implies, since ${\rm Im} \,  \mathcal{N}_{\Lambda \Sigma}$ must be invertible for a consistent quantization of supergravity,
 that
\be
f^\Lambda _i=0\, .
\ee
In such case it follows from eq. (\ref{g}) that
\be \qquad g_{i\bar j}=0\, .
\ee
Also for constant ${\cal N}_{\Lambda \Sigma}$ eq. (\ref{C}) implies that
$
C_{ijk}=0
$.
The vanishing metric of the \K\, geometry $g_{i\bar j}=0$ means that there are no vector multiplet scalars. In fact, $\partial_i   \, {\cal N}_{\Lambda \Sigma} =0$ in $N>1$ is only possible if there are no scalars, except for hypers in $N=2$ case.  For example, $N=3$ models of the type defined in \cite{Ferrara:1989nm,Frey:2002hf} can be reduced to $N=2$ models with vector multiplets with $C_{ijk}=0$ and hypermultiplets. When further reduced to $N=1$ these models have only minimal kinetic terms \cite{Andrianopoli:2002vq}.

We conclude therefore that eq. (\ref{NM}) is compulsory for a consistent $N=2$ special geometry.  We remind that in this case we only used properties of special geometry which encompass both symmetric and non-symmetric special \K\, manifolds.

$N=2$ non-symmetric space models of special geometry also universally require a non-minimal vector coupling. These couplings remain non-minimal and scalar dependent, when truncated to $N=1$, provided that $C_{ijk}\neq 0$. So, special geometry with a non-vanishing $C_{ijk}$ tensor can be regarded as a generalization of groups of type E7 for non-symmetric special manifolds, see the Appendix for the details.

It is particularly interesting that the simplest cases of non-symmetric spaces with a non-vanishing $C_{ijk}$ correspond to the Calabi-Yau special geometries, as they come in type II string theory compactifications \cite{Cecotti:1988qn}. The related c-map hypergeometry is also non-symmetric (although it is  an Einstein space). It may also be useful to remind here that the examples of $10^{500}$  string landscape vacua   \cite{Denef:2004dm}
 are based on such Calabi-Yau threefold with the non-vanishing triple intersection numbers $C_{ijk}$.

\subsection{All kinetic terms in $N>1$}
We have focused our attention to the fact that the vector kinetic terms are non-minimal since we are interested in cosmological applications of the vertices which allow the decay of the inflaton and unwanted relics into particles of the standard model. However, we may add here that all kinetic terms in $N>1$ supergravities are ``non-minimal'' which means that  they depend on scalars. For supergravity manifolds of $N>2$ this is obvious because they are Einstein spaces  with (negative) constant curvature.

For $N=2$ theories vector multiplets are coupled to special geometry space which is never flat, while hypermultiplets are described  by quaternionic geometries which are generically not symmetric and belong to Einstein spaces with constant negative curvature \cite{Bagger:1983tt}. This means that the kinetic terms for scalars in all cases of $N=2$ are non-minimal, however, the vector kinetic matrix depends only on scalars from the vector multiplets and does not depend on hypers.

\section{Consistent truncations of $N>1$ to $N=1$}
A detailed study of the consistent truncation of $N>1$ supergravity to an $N=1$ was performed in
\cite{Andrianopoli:2001zh}. Here we would like to  ask whether a theory with $N>1$ can be consistently truncated to an $N=1$ supergravity with minimally coupled vectors. This is a very strong constraint since it would require not only the matrix $\mathcal{N}$
to become holomorphic, but in fact constant. For symmetric spaces this is only possible in one particular case if we consider the $CP^n$ models with $C_{ijk}=0$, keep $n_c$ chiral multiplets and $n_v=n-n_c$ vector multiplets. These models may also include some hypermultiplets decoupled from vector multiplets.
Then the reduced theory will have a constant $\mathcal{N}$
 matrix.
 
 For $N\geq 4$ consistent truncations to $N=2$ exist which include, beyond vector multiplets, also hypermultiplets (see f. e. Table 1 in \cite{Ferrara:2007pc}). Then further reduced to $N=1$ hyperscalars will be minimally coupled to vectors. However, the scalar sector coming from vectors will not be minimally coupled, since in general $C_{ijk}\neq 0$. There is only a universal truncation to $N=1$ which is minimally coupled which is the axion-dilaton $N=2$ multiplet further truncated to a $N=1$ vector multiplet. We observe that this  is always possible because this corresponds to the $N=1$ truncation of a decoupled space for a ${SL(2, \mathbb{R})\over U(1)}\times {SO(2,n)\over SO(2)\times SO(n)} $ manifold when we set $n=0$.
 
 We then conclude that minimally coupled scalars in $N=1$ truncation from higher $N$ can only be obtained if, at the level of $N=2$ the scalars come from hypermultiplets or from $CP^n$ submanifolds of the $N=2$ manifolds (the axion-dilaton N=2 multiplet corresponds to the $CP^{1}$ case in which case no scalar is left from this sector).

 Following \cite{Andrianopoli:2001zh} the ${\cal N}_{\Lambda \Sigma}$ matrix in $N=2$ is
\be
{\cal N}_{\Lambda \Sigma}= \bar F_{\Lambda \Sigma}-2i \bar T_\Lambda \bar T_\Sigma(L^Z {\rm Im} F_{ZW} L^W)\, ,
\label{calN}\ee
where $F_{ZW}$ is the holomorphic second derivative of the prepotential,  $L^Z$ is the relevant part of the covariantly holomorphic section and $T_\Lambda$ are the projectors on the graviphoton. We now redefine the indices so that the original $\Lambda$ taking values  $ 0,1,..., n$ is split into an index $\alpha$ taking values $1, ..., n_v$ for $N=1$ vector multiplets and an index $X$ taking values in $0,1, ..., n_c$. Here $n=n_c+n_v$. Note that the $N=1$ vector index $\alpha$ does not include $0$.

A consistent truncation to $N=1$ from $N=2$ models with $C_{ijk}=0$ means that one starts with the quadratic holomorphic prepotential $F(X)$. In such case one finds from (\ref{calN}) that
\be
{\cal N}_{\alpha \beta}= \bar F_{\alpha \beta}=-i \delta_{\alpha \beta}
\ee
and ${\rm Im} {\cal N}<0$ as it should be. Here the relation between the $N=2$ and $N=1$ 	 kinetic matrix is $\bar F_{\alpha \beta} ={\cal N}_{\alpha\beta}= -{i\over 4}\bar f_{\alpha \beta} (\bar z)$ which agrees with \cite{Cremmer:1982en}.																												 

Generic models derived from higher dimensions \footnote{With exception of the string theory derivation of $N=3$ supergravity  model with $U(3,n)$ duality group  based on  a ${T^6\over Z^2}$ orientifold   \cite{Frey:2002hf}.} have $C_{ijk}\neq 0$. Therefore the non-minimal vector coupling $\mathcal{N}_{\Lambda\Sigma}$  reduced to $N=1$ will not be constant, in general. As an example one can consider the  effective action of type $N = 1$  Calabi-Yau orientifolds derived from the compactification of  string theory in \cite{Grimm:2004uq}.
 This is a   generic case of  superstring inspired $N=1$ supergravities which arise from  compactified  Calabi-Yau  orientifolds. They produce the cubic coupling associated with the holomorphic vector coupling
\be
\overline {\cal N}_{\alpha \beta}(z)= d_{\alpha \beta p} \, z^p\, .
\ee
Here  as above $\alpha, \beta = 1, ..., n_v$ and $p=1, ..., n_c$ and $d_{\alpha \beta p}$ codifies  the Calabi-Yau  threefolds intersection numbers. The kinetic holomorphic vector coupling matrix is linear in the scalars of a chiral multiplets $z^p=a^p +i \varphi^p$ and is proportional to
\begin{eqnarray}
& d_{\alpha \beta p} \, \varphi^p F^{\alpha}_{\mu \nu}F^{\beta \mu\nu}
+ d_{\alpha \beta p} \, a^p F^{\alpha}_{\mu \nu}\tilde F^{\beta \mu\nu}\, .
\label{nonminimal}
 \end{eqnarray}
Thus, if one of the scalars in the chiral multiplet $z^p$ or their combination is an inflaton field or any other heavy scalar we would like to get rid of, it has a cubic vertex and a route of decay into two vectors. In the same models there is a Pauli interaction of the following form, see eq. (\ref{Pauli})
\be
d_{\alpha \beta p} \, \bar \chi^p \sigma_{\mu\nu} \lambda^{\alpha} F^{\beta \mu\nu}\, .
\ee
If the fermion partner of the inflaton or  any other heavy fermion was created at the non-perturbative stage of preheating, it will also decay into a gaugino and a vector field at a later stage via the compulsory Pauli interaction.

\section{Discussion}

The minimal $N=1$ supergravity  seems to be the simplest one and described in \cite{Wess:1992cp} with the minimal vector coupling (\ref{minimal}). It may be viewed, however,  as an incomplete theory, when the creation of matter in the Universe after inflation is studied. In this paper we have argued that any version of $N=1$ supergravity inspired by either supergravities with more supersymmetries (extended $N>1$ supergravities), or originating from higher dimensions and compactifications of superstring theory, must have  a non-minimal scalar dependent vector couplings, as well as Pauli couplings. In non-supersymmetric theories the coupling $a F\tilde F$ of vectors with axion field  $a(x)$ was studied intensely in QFT and in applications to cosmology, see for example \cite{Barnaby:2011qe} and references therein.
The couplings of vectors to moduli fields  $e^\phi  F F$
were also studied in various situations in QFT and cosmology.

In this paper we proved that in $N=1$ supergravity obtained via truncation from extended $N>1$ supergravities, or originating from higher dimensions and compactifications of superstring theory, the interactions between vectors and scalars and the corresponding Pauli interactions are compulsory (with exception of $U(p,n)$ models and hyperscalars), namely $N=1$ action includes the following interactions:
\be
{\rm Im}\,  {\cal N}_{\Lambda \Sigma}(\phi)\,
F_{\mu\nu}^\Lambda F^{\mu\nu \Sigma}+  {\rm Re} \, {\cal N}_{\Lambda \Sigma}(\phi) \,   F_{\mu\nu}^\Lambda \tilde F^{\mu\nu \Sigma } +{ \partial   \, {\cal N}_{\Lambda \Sigma}\over \partial \phi^p} \, \bar \chi ^p \sigma_{\mu\nu} \lambda ^\Lambda F^{\Sigma \mu\nu}+...\, .
\label{v}\ee
When the kinetic function ${\cal N}_{\Lambda \Sigma}(\phi) $ is linear in scalars
\be
\overline {\cal N}_{\Lambda \Sigma}(\phi) =d_{\Lambda \Sigma p} z^p
\ee
and $d_{\Lambda \Sigma p}$ are some constants (which have a simple interpretation in Calabi-Yau compactifications or special geometry in $N=2$),  the theory has a  cubic coupling between scalars and vectors. If the scalar is an inflaton or some heavy scalar it has a simple route of decay via creation of vector particles, which in turn couple to other standard model particles and create the whole matter in the universe.
Some unwanted  fermions, partners of inflaton or of the heavy scalars, will also decay via the cubic Pauli coupling.

A remarkable feature of $N>1$ supergravities is the fact that the non-minimal vector couplings and Pauli couplings are compulsory and universal (with exception of models in $N=2$ which have only hyper scalars). One quick look at Tables 1  and 2, which give the list of duality symmetries $G$ of all  symmetric space $G/H$ supergravities, shows how different they are and how distinct are their duality groups $G$. The non-symmetric spaces of $N=2$ special geometry have enormous amount of choices associated with the non-vanishing 3-forms $C_{ijk}$.  The proof of non-minimal vector couplings for all $N>1$ symmetric as well as non-symmetric  supergravities is based on the universal feature of all of these models:  they are based on duality groups of type E7.  The proof follows from the definition of these groups. 

Exceptional cases of $U(p,n)$, $p=3,1$  groups, which are not of type E7, also require a non-minimal vector coupling in $N>1$ models,  the proof being based on the non-compactness of these groups for $n\neq 0$ and negative definiteness of the kinetic term for vectors.

 The vector couplings remain non-minimal, scalar dependent, when truncated to $N=1$ for all models of type E7, which form a majority of extended supergravity models. These couplings become minimal, scalar independent,  in the exceptional case $U(p,n)$, $p=3,1$, with  vanishing 3-forms $C_{ijk}=0$ in $N=2$ case,  or when all scalars originate from the hypers.  
 
 This universality of a scalar dependent vector coupling may have a cosmological significance:  it suggests that creation of matter in the universe via non-minimal vector and Pauli couplings may be a dominant factor in theories inspired by superstring theory and extended supergravity.

\

\

{\Large {\bf Acknowledgments}}

It is a pleasure to thank  R. Bond, J. Broedel, J.J. Carrasco, A. Linde,  L. McAllister, K. Olive and E. Silverstein for stimulating conversations.    The work of
S.F.~has been supported by the ERC Advanced Grant no. 226455, ÒSupersymmetry, Quantum
Gravity and Gauge FieldsÓ (SUPERFIELDS), and in part by DOE Grant DE-FG03-91ER40662.
The work of R.K. was supported by the Stanford Institute of Theoretical Physics and NSF grant 0756174.

\appendix

\section{Special Geometry and Groups of Type E7}
The contact of special geometry with groups of type E7 can be made by introducing  a quartic  functional of central charge $Z$ and matter charges $Z_i= D_i Z$, as shown in  \cite{Cerchiai:2009pi}
\be
{\cal I} (\phi, Q)=(Z\bar Z- Z_i \bar Z^i)^2 + {2\over 3} i (Z N_3 (\bar Z) - \bar Z \bar N_3(Z))- g^{i\bar i} C_{ijk} C_{\bar i \bar j \bar k} \bar Z^j \bar Z^k  Z^{\bar j} Z^{\bar k}\, .
\ee	
The quartic invariant ${\cal I} (\phi, Q)$ is actually the coordinate covariant expression of the ``h'' function
introduced in eq. (2.31) of 
\cite{deWit:1992wf} if one replaces (half of) the quaternionic coordinates with the dyonic charge vector $Q$, in the c-map construction of \cite{Cecotti:1988qn}.

This quartic invariant has the property that at the attractor points
\be
2\bar Z Z_i+i C_{ijk}\bar Z^j \bar Z^k=0
\ee
it takes the following value
\be
{\cal I} (\phi_H, Q)=V_H^2-{32\over 3} |Z|^2(Z_i \bar Z^i)_H \ ,
\ee	
where $V_H$ is the black hole potential (at the attractor point) at the black hole horizon
\cite{Ferrara:2011gv}.
As a result, either
\be
{\cal I} (\phi_H, Q)= {\cal I}_4 (Q)
\ee
for symmetric special geometries of E7 type, or
\be
{\cal I} (\phi_H, Q)= ({\cal I}_2 (Q))^2
\ee
for symmetric geometries with $C_{ijk}=0$, i.e. ${U(1,n)\over U(1) \otimes U(n)}$  $CP^n$ models.
Therefore, as stated in the text, in $N=2$ groups of type E7 require $C_{ijk}\neq 0$, which is the condition for the existence of the ``primitive quartic invariant''.

For non-symmetric spaces the  quartic functional exists (as shown in eq. (4.2) of \cite{Ferrara:2011gv}) however, it is not invariant as it is scalar dependent. Note that the case $C_{ijk}=0$ is the only case which can give a consistent truncation to the minimally coupled $N=1$  theories.

\end{document}